\documentclass[journal]{IEEEtran}

\usepackage{setspace,amsmath,latexsym,cite,amssymb,epsfig,amsfonts}
\usepackage{url,cite}
\usepackage{graphicx}
\usepackage{psfrag}
\usepackage{footmisc}
\usepackage{multirow}
\usepackage{color}

\graphicspath{{.}{../eps/}{./images/}{./images/result/}}
\usepackage{amssymb}
\usepackage{amsthm}

\usepackage{cite}
\usepackage{mathtools, cuted}
\usepackage[font=small, labelsep=period]{caption}
\usepackage{makecell}
\usepackage{stfloats}
\usepackage{subcaption}
\usepackage{changepage}
\usepackage{balance}

\allowdisplaybreaks[4]

\author{Chong Huang, \IEEEmembership{Member, IEEE}, Gaojie Chen, \IEEEmembership{Senior Member, IEEE}, Pei Xiao, \IEEEmembership{Senior Member, IEEE},\\Zhu Han, \IEEEmembership{Fellow, IEEE}, and Rahim Tafazolli, \IEEEmembership{Fellow, IEEE}
\thanks{C. Huang, P. Xiao and R. Tafazolli are with 5GIC \& 6GIC, Institute for Communication Systems (ICS), University of Surrey, Guildford, GU2 7XH, United Kingdom, Email: \{chong.huang, p.xiao, r.tafazolli\}@surrey.ac.uk.}
\thanks{G. Chen is with School of Flexible Electronics (SoFE) \& State Key Laboratory of Optoelectronic Materials and Technologies, Sun Yat-sen University, Guangdong, China, Email: chengj235@mail.sysu.edu.cn.}
\thanks{Z. Han is with the Department of Electrical and Computer Engineering at the University of Houston, Houston, TX 77004 USA, and also with the Department of Computer Science and Engineering, Kyung Hee University, Seoul, South Korea, 446-701. Email: hanzhu22@gmail.com.}
}

\begin{document}

\title{\Huge Large Artificial Intelligence Models for Future Wireless Communications}

\maketitle
{
\begin{abstract}
The anticipated integration of large artificial intelligence (AI) models with wireless communications is estimated to usher a transformative wave in the forthcoming information age. As wireless networks grow in complexity, the traditional methodologies employed for optimization and management face increasingly challenges. Large AI models have extensive parameter spaces and enhanced learning capabilities and can offer innovative solutions to these challenges. They are also capable of learning, adapting and optimizing in real-time. We introduce the potential and challenges of integrating large AI models into wireless communications, highlighting existing AI-driven applications and inherent challenges for future large AI models. In this paper, we propose the architecture of large AI models for future wireless communications, introduce their advantages in data analysis, resource allocation and real-time adaptation, discuss the potential challenges and corresponding solutions of energy, architecture design, privacy, security, ethical and regulatory. In addition, we explore the potential future directions of large AI models in wireless communications, laying the groundwork for forthcoming research in this area.
\end{abstract}
}
\IEEEpeerreviewmaketitle
\section{Introduction}\label{sec:1}
Wireless communications have evolved through various stages, leading to the current fifth-generation (5G) era and looking ahead to the forthcoming sixth-generation (6G), each phase has introduced notable improvements to meet the increasing demands of societal progression for enhanced wireless digital connectivity. However, with the development of wireless networks, there is a perceptible trend towards more complex and large-scale communication networks, this leads to the high complexity of network optimization and management. As the Internet of Things (IoT), smart cities and intelligent vehicles develop in modern society, functional demands on wireless networks are set to grow exponentially. Considering the increasing expectations of wireless communications, traditional methods face increasing challenges for designing future wireless networks.

In the past years, artificial intelligence (AI) has experienced profound advancements. Recently, a notable development in current AI research is the emergence of large AI models. The large AI models are characterized by a large amount of parameters and expensive training cost, and have a wide spectrum of various applications, such as natural language processing and computer vision, etc. Leading this wave are transformer-based large language architectures, including OpenAI's GPT series \cite{NEURIPS2020}, Google's LaMDA \cite{NEURIPS2022lamd}, and Meta's LLaMA \cite{min2021recent}. Given the considerable potential of these large AI models, their expertise is not limited to natural language processing. The large AI models can also utilize training datasets from diverse tasks to serve various domains, encompassing image recognition, chess strategizing and mathematical reasoning.

Considering the powerful adaptability and dynamic learning ability of AI, its integration with the wireless communication systems has significant potential in the future. Thus, with the advent of 5G, many studies have already pivoted their focus to the fusion of 6G and AI\cite{8970161}. Furthermore, the application of edge intelligence in wireless networks has been investigated to facilitate communication-computation separation, and enhance the performance in distributed wireless systems \cite{8808168}. The integration of AI with 5G and beyond-5G (B5G) has also been studied to explore the potential of introducing AI algorithms in channel measurement, network optimization, and many other areas of network management \cite{9023918}. Moreover, considering the key role of latency in wireless communication, the application of AI in ultra-reliable low-latency communications (URLLC) was studied to optimize information transmission delays \cite{9023932}.

Although the integration of AI into wireless communications is not a novel topic, utilizing large AI models for wireless networks remains an under-developed area. The scale and complexity of large AI models present unprecedented possibilities for the design of future wireless networks. These large AI models have hundreds of billions to trillions of parameters, they are not only information processing systems but also unified management centers capable of dynamic learning, predictive analytics and self decision-making. Utilizing these AI models, especially the large AI models, promises capabilities like dynamic real-time resource allocation, network demand forecasting, and transmission efficiency enhancement. Recently, China Telecom has released and deployed the first large AI model Qiming for the management of communication networks, this large AI model is trained on a huge dataset and integrates an array of sub-AI models including AI models for network management, algorithm development, computational resource allocation and service optimization. As the demarcation between communication and computation becomes increasingly nebulous, these advanced AI solutions are estimated to play a key role in shaping the efficiency, speed, and versatility of the next generation of wireless systems.

However, the evolution of integrating wireless communication and AI has not been exempt from challenges. The huge scale and high complexity of large AI models raise concerns about computational efficiency, real-time data processing, and energy consumption in wireless networks. Furthermore, the integration of large AI and wireless communications requires to design new underlying network architectures, protocols and security rules. Nonetheless, the prospective benefits of the integration of large AI and wireless communication networks can significantly outpace their challenges and costs. This AI-driven evolution not only brings improvements in transmission rate and efficiency, but also catalyzes innovations in services and applications.

To fully realize the potential of large AI in wireless communications, a comprehensive and multidimensional approach is required in future wireless systems. Technologically, pioneering works in areas such as model compression, edge computing, and network architectural designs are very important to realize the AI-based wireless communications, these key techniques can help enhance the efficiency of introducing large AI models in future wireless networks. Besides, in terms of ethical and security, data privacy and security concerns present key challenges for future AI-driven wireless systems. Considering that large AI models can learn everything very fast, it is necessary to define security standards to ensure AI-driven decisions in wireless communications.

In conclusion, the incorporation of AI into wireless communication signifies the frontier of the next information age, it heralds an era of Internet of Everything (IoE). Although current research in wireless communications focuses on the fusion of singular systems with AI, including energy resource management \cite{9679392}, network defense algorithm design \cite{9706362}, multi-access edge computing (MEC) resource allocation \cite{10015857}, satellite communications \cite{9420293} and semantic communications \cite{10038754}, the exploration and application of more large AI architectures in this area remain uncharted. This paper will investigate the confluence of large AI models and future wireless communications, offers insights into their potential, challenges and future directions for a large AI-driven wireless communication.

\section{Large AI Models in Wireless Communications}\label{sec:2}
The rapid development of AI has brought capabilities that were once deemed unattainable. In wireless communications, the prowess of these AI models are utilized to predict, optimize, and revolutionize various facets of the communication process. In this section, we will discuss the advantages that AI offers to wireless communication and explore how large AI models utilize these benefits to further enhance performance and even bring transformative changes.

\subsection{Typical AI-Driven Applications in Wireless Communications}
\begin{figure}[t!]
  \centering
  \centerline{\includegraphics[scale=0.45]{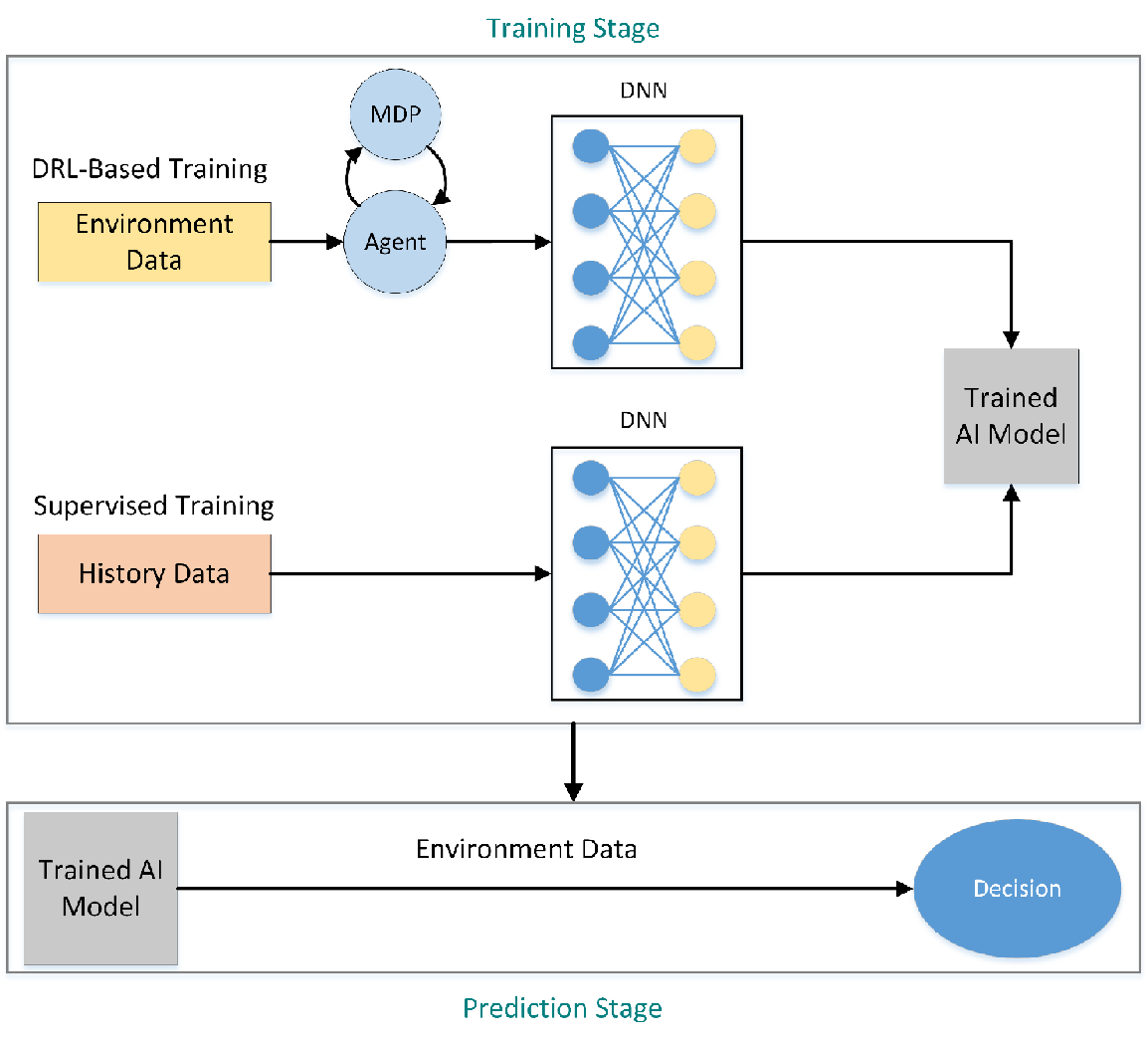}}
 \caption{ Traditional AI model for wireless communication tasks.} \label{fig:sm1}
\end{figure}

Integrating AI into wireless communication has emerged as an important strategy to address the high complexity and demands of modern and future communication networks. AI shows substantial potential in enhancing various facets of wireless communications, including efficiency, reliability and security. As illustrated in Fig. \ref{fig:sm1}, AI training in wireless communications mainly includes supervised deep learning and deep reinforcement learning (DRL).

Supervised learning, often considered as the cornerstone of deep learning in AI, it is the paradigm where virtually all deep networks are trained. It trains AI models on labeled datasets to approximate functions that correlate input and output data seamlessly. In contrast, DRL, an integration of deep learning and reinforcement learning, has emerged as the solution in dynamic and uncertainty environments, because labeled data is hard to be obtained in these environments. DRL employs Markov-decision-processes (MDP) to adapt its behavior to train optimal models in such complex environments.

The following describe some prevalent AI-based applications in wireless communication networks, underscoring their contributory roles in enhancing efficiency and optimizing resource allocations.

\subsubsection{Edge Computing}
Edge computing, emphasizing data processing closer to the data source rather than centralized cloud-based systems, is envisaged to play a key role in future wireless communications. By deploying AI at the edge, operators can make predictions based on user requirements, local network conditions, and device behavior in real-time, to enhance the network efficiency and responsiveness \cite{10015857}.

\subsubsection{Semantic Communications}
Semantic communications encompass data transmission predicated on its overarching meaning or high-level information rather than mere raw data transmissions. Through predicting the semantics of data packets, generative AI models can be employed to reconstruct the original content \cite{10038754}. Therefore, by only transmitting key information, semantic communications effectuate a breakthrough in Shannon capacity, and are particularly advantageous in network congestion or bandwidth-limited scenarios.

\begin{figure*}[t!]
\begin{adjustwidth}{0em}{0em}
\begin{subfigure}[b]{.45\textwidth}
  \centering
  \includegraphics[width=1\linewidth]{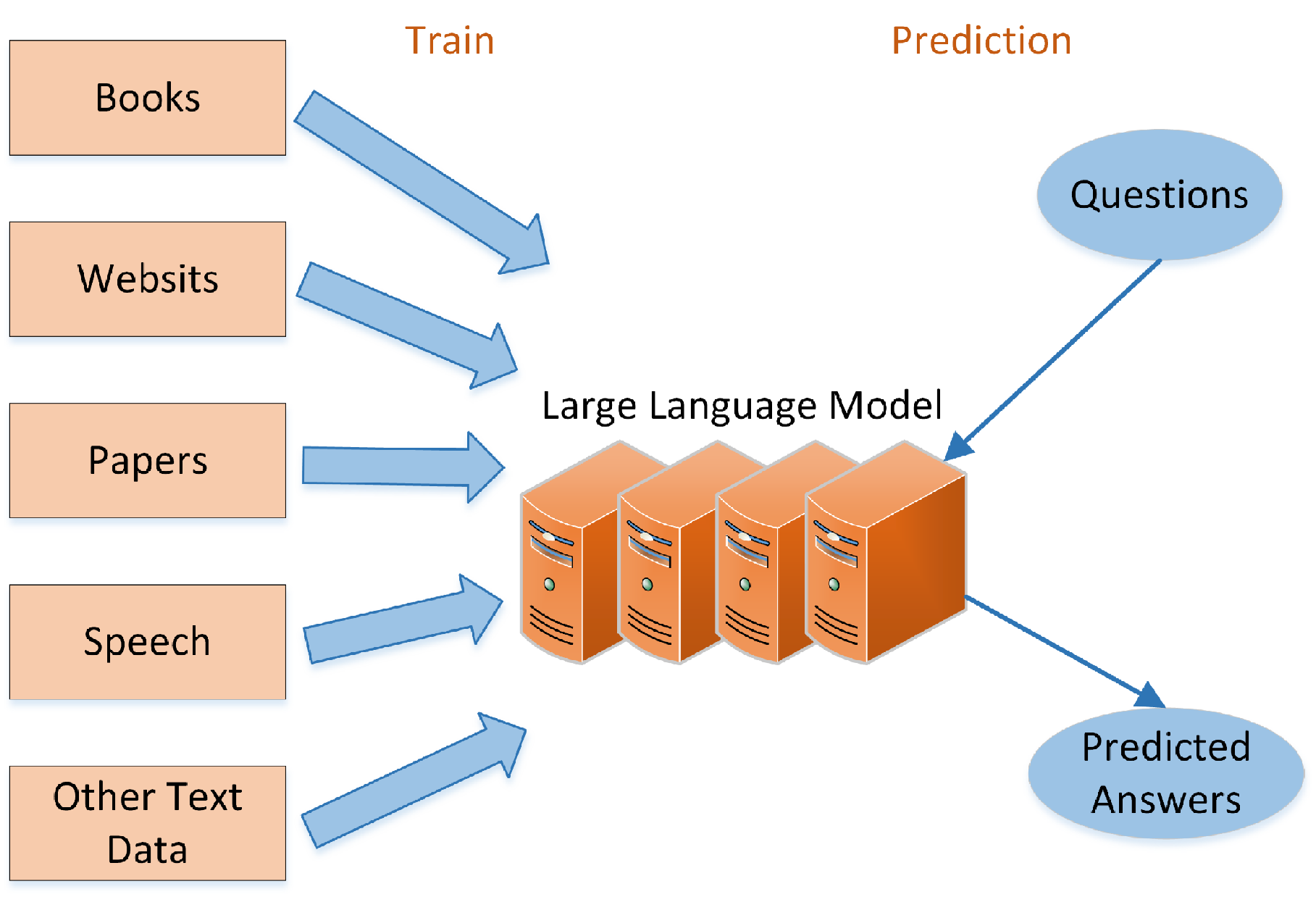}
  \caption{Large language models. }
  \label{fig:sm2_1}
\end{subfigure}\hspace{10mm}
\begin{subfigure}[b]{.45\textwidth}
  \centering
  \includegraphics[width=1\linewidth]{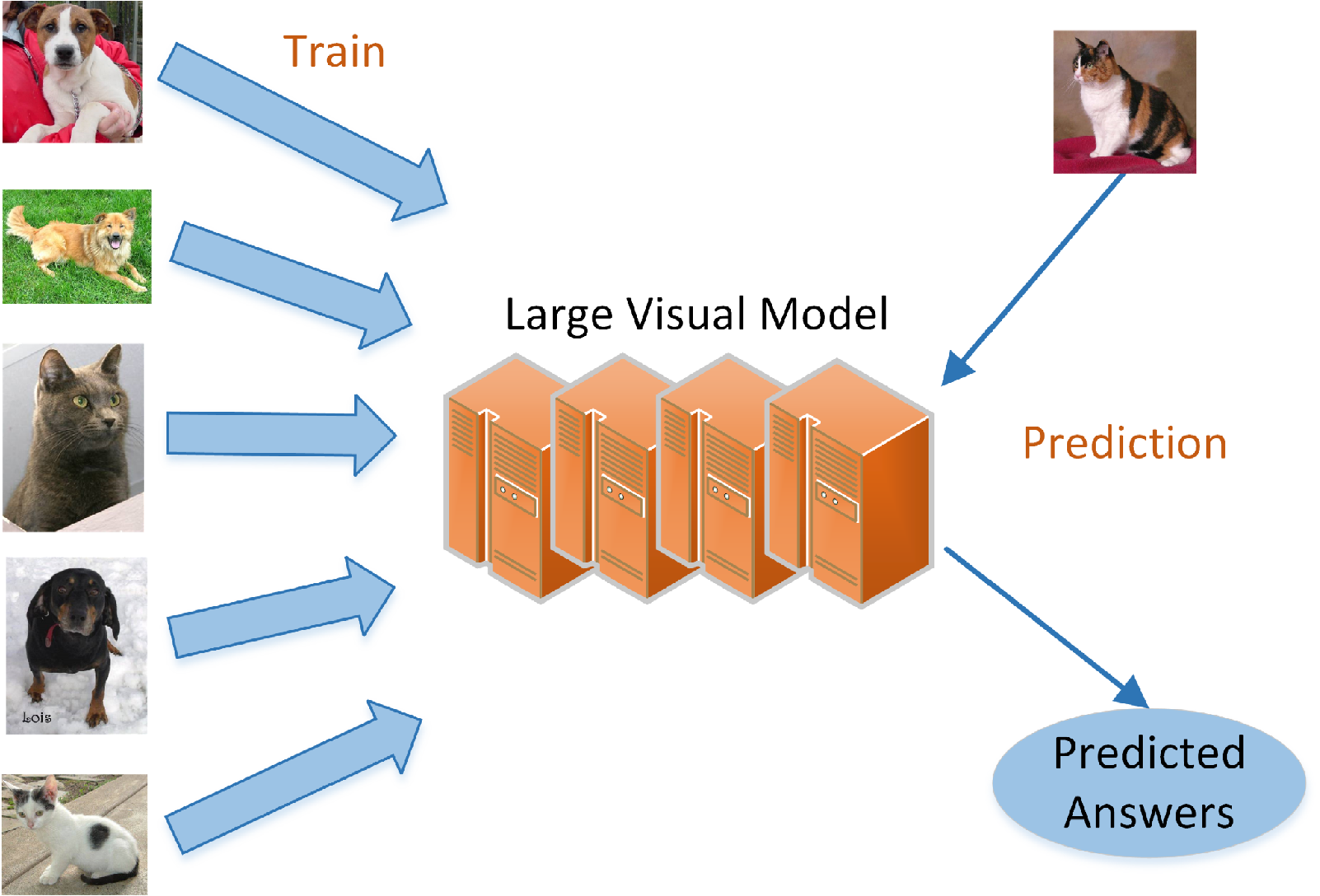}
  \caption{Large visual models.}
  \label{fig:sm2_2}
\end{subfigure}
\end{adjustwidth}
\caption{Existing large AI models.}
\label{fig:sm2}
\end{figure*}

\begin{figure}[t!]
  \centering
  \centerline{\includegraphics[scale=0.27]{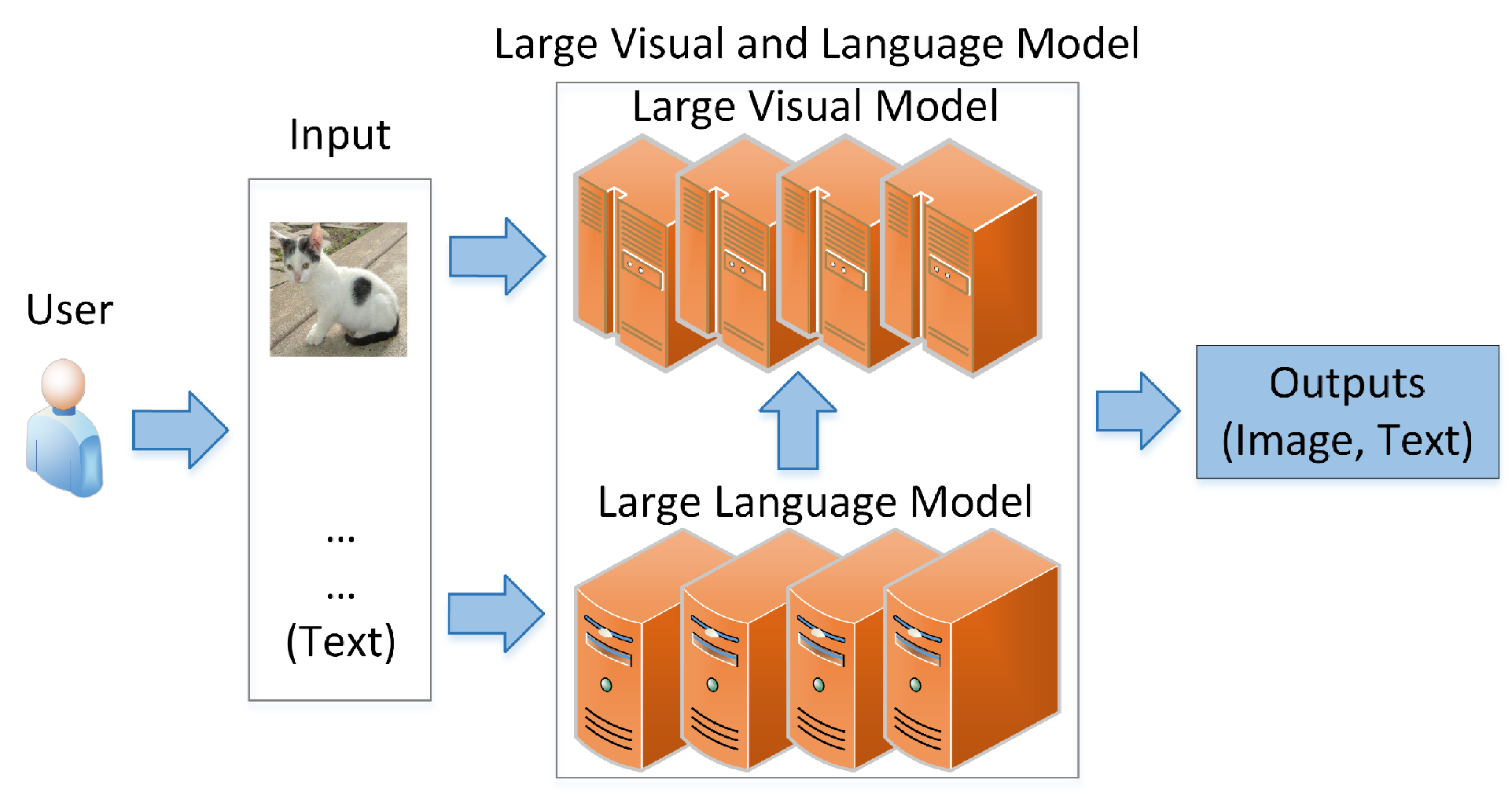}}
 \caption{ Large language and visual AI models.} \label{fig:sm2_3}
\end{figure}

\subsubsection{Security Threat Prediction}
Considering the growing complexity in wireless communications, they also become more vulnerable to security threats. AI models can estimate potential vulnerabilities by learning from past threat and attack pattern datasets \cite{9706362}. This proactive security paradigm seeks to preemptively mitigate threats, safeguarding both communication networks and users.

\subsubsection{Energy Efficiency and Green Communications}
With sustainability gaining precedence, reducing energy consumption in wireless communications is imperative. AI-driven green communications are utilized to optimize network performance for the goal of minimizing energy consumption and environmental impact \cite{9679392}. AI algorithms are designed to enhance energy efficiency at various levels of wireless communications, from optimizing signal processing and reducing transmission power to enhancing the energy efficiency of base stations, users, data centers and the entire network. Therefore, the integration of renewable energy sources, energy harvesting techniques and energy-efficient hardware in AI-driven wireless networks is a significant area towards future green communications.

\subsubsection{Satellite Communications}
With the advent of mega-constellations like Starlink, satellite communications are becoming an integral facet of global wireless connectivity. AI models consider satellite positions, states, user locations, mobility patterns and network conditions to predict optimal satellite handovers, beamforming directions and ground station allocations. This ensures high-quality communications even in traditionally inaccessible or remote areas.

\subsubsection{User Behavior and Demand Forecasting}
Comprehending and predicting user behavior is important for optimal wireless network resource allocation. AI models can predict user access peak times, locations, and corresponding services by analyzing historical user data traffic, movement patterns, and device usage tendencies. This enables network providers to allocate resources more efficiently to ensure high-quality service all the time.

\subsection{Large AI Models in Wireless Communications}
Recently, many natural language models have emerged and give rise to multi-function large AI models such as ChatGPT \cite{NEURIPS2020}. As illustrated in Fig. \ref{fig:sm2}, large language models are trained on huge text datasets including books, speeches, web pages, and academic papers, and these models are adept at understanding user's text inputs and generating appropriate responses \cite{10648594}. In addition, the prominent models in current works are not limited to natural language processing, they have also expanded into large visual models, as depicted in Fig. \ref{fig:sm2_1}. These large visual models are trained on amount of images to extract features and information, subsequently enabling image classification and semantic recognition \cite{10558819}.

\begin{figure}[t!]
  \centering
  \centerline{\includegraphics[scale=0.48]{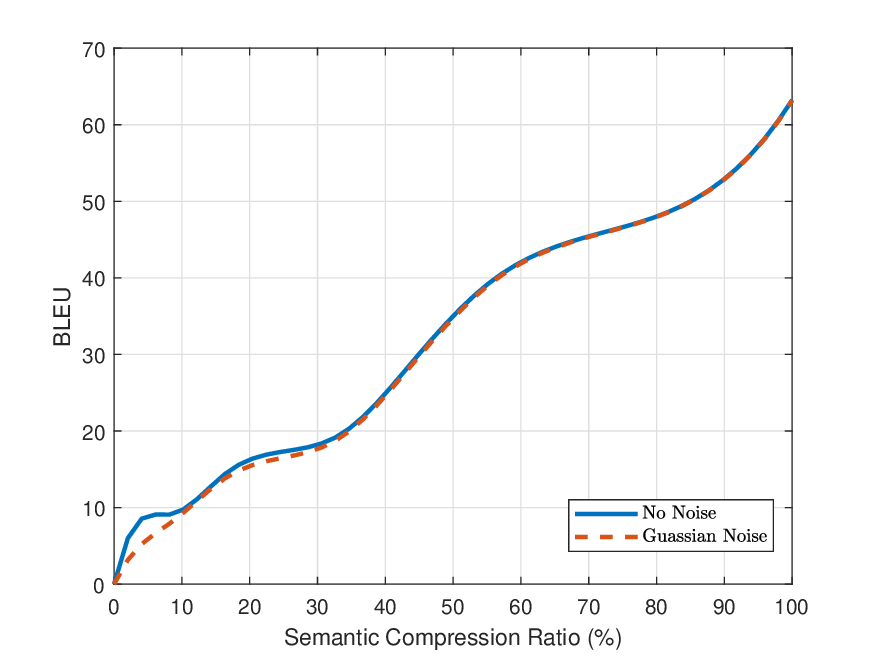}}
 \caption{Large AI model in text semantic communications.} \label{fig:r1}
\end{figure}

\begin{figure}[t!]
  \centering
  \centerline{\includegraphics[scale=0.48]{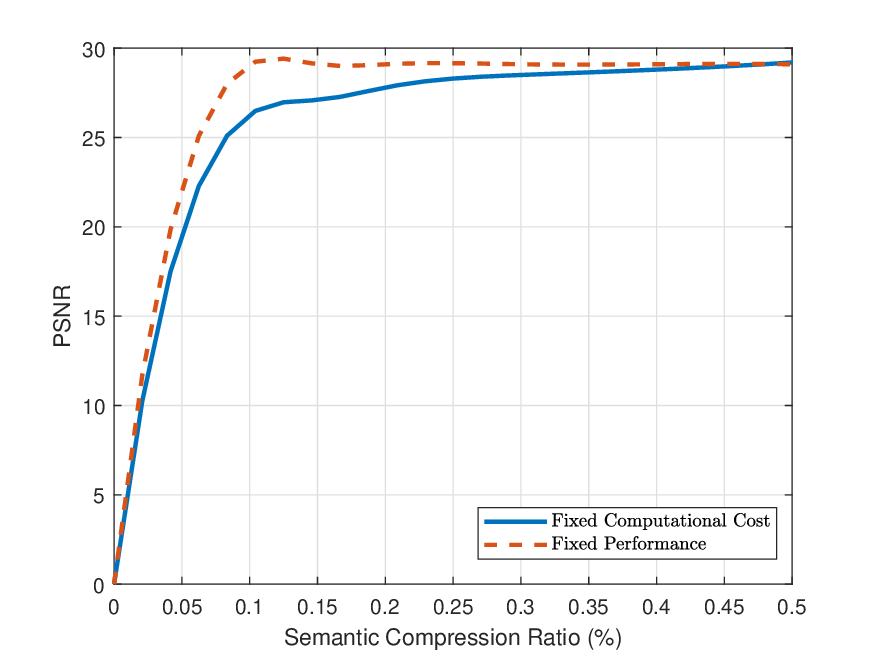}}
 \caption{Large AI model in image semantic communications.} \label{fig:r2}
\end{figure}

In this study, we investigate the performance of using large AI models for text and image semantic communications in wireless networks. As shown in Fig. \ref{fig:r1}, with an increase in the compression ratio (i.e., the size of compressed semantic information becomes larger), the bilingual evaluation understudy (BLEU) score for the restoration of semantic information at the receiver also increases, allowing more semantic content to counteract distortions during semantic transmissions. Moreover, semantic noise has a significant impact on the accuracy of semantic transmissions, especially under high compression demands. Besides, as shown in Fig. \ref{fig:r2}, semantic compression of images is more effective than that of text, significantly reducing transmission bandwidth usage because semantic parsing of natural language is more challenging than that of images. Considering that the image transmission consumes more spectrum resources than text in communication systems, semantic transmissions of images will play a key role in future wireless communication systems. Therefore, integrating semantic information from both images and text to enhance the generating ability at the receiver in semantic communications is a key research direction in current semantic communications.


On the other hand, as previously highlighted, the potential of large AI models are not limited to only language or visual processing. As shown in Fig. \ref{fig:sm2_3}, a combined large AI model named VisualGPT \cite{wu2023visual} which integrates both visual and language comprehension to show the multi-modal potential of large AI models. Such AI models can output language, images and decisions, making them adept at providing responses including language outputs, image representations, optimized solutions, and parameters upon receiving both visual and text inputs.

\begin{figure*}[t!]
  \centering
  \centerline{\includegraphics[scale=0.48]{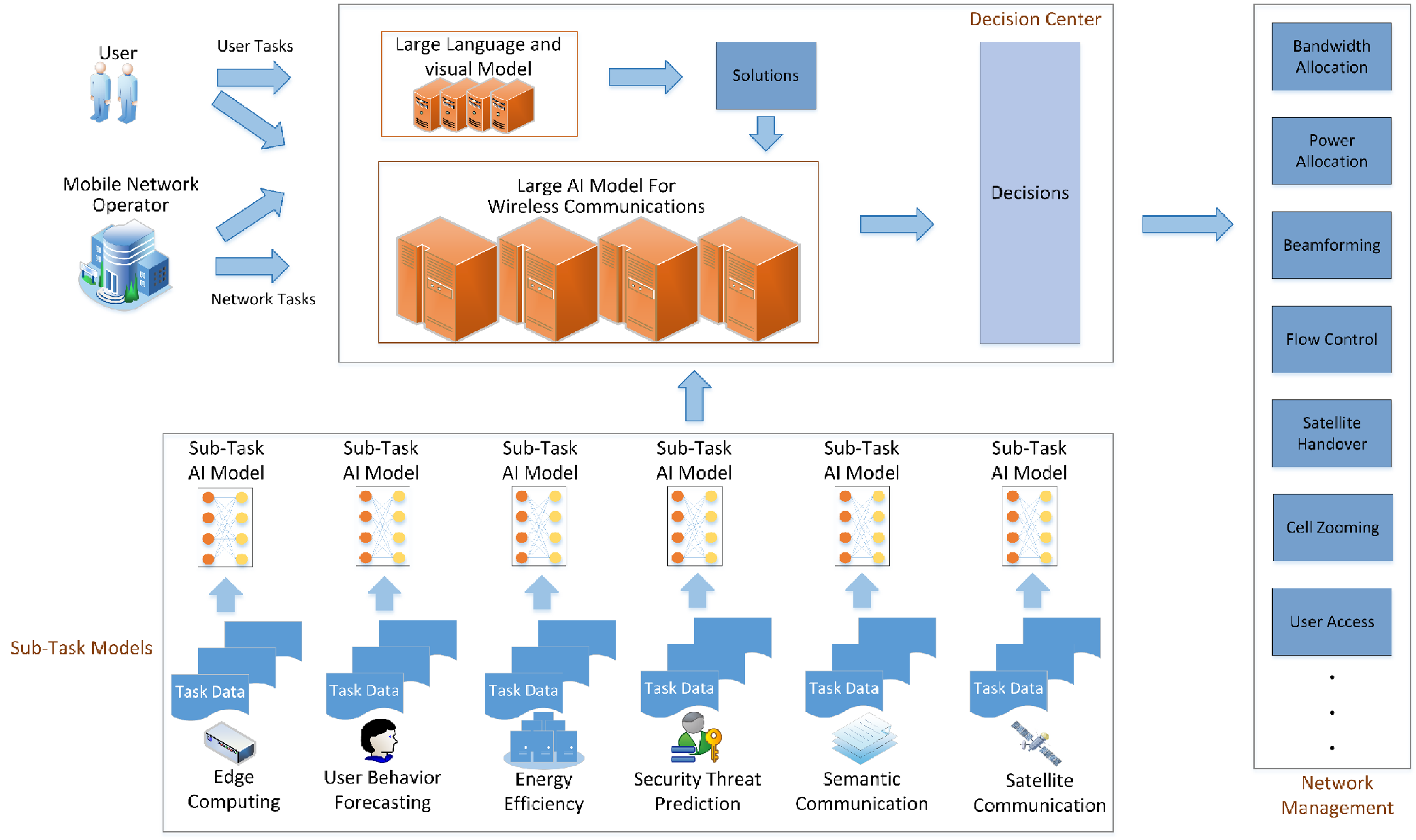}}
 \caption{Large AI model in future wireless communications.} \label{fig:sm3}
\end{figure*}

Therefore, the future large AI models can be multi-modal. Existing and emerging AI models, exemplified by Qiming, can be combined and trained according to the specific needs of wireless communication networks, improving the decision-making hub that dynamically and autonomously optimizes decisions and boosting energy efficiency in wireless communications. As illustrated in Fig. \ref{fig:sm3}, future large AI models in wireless communications are capable of consolidating various sub-task AI models, each specialized for specific management area such as resource optimization in edge computing, content generation in semantic communications and intelligent scheduling in satellite communications. Consequently, future large AI models in wireless communications will essentially serve as a conglomerate of AI clusters, making decisions based on the information derived from individual sub-tasks. Moreover, future wireless communications are estimated to embed large visual and language models into the large AI models to enhance the understanding of user and operators requirements. In current wireless networks, users are reliant on software interfaces to send communication requirements and commands. However, with the assistance of large visual and language models, users can send instantaneous communication requirements by voice or text. The large AI models are equipped with natural language processing capabilities to interpret these text and voice inputs, subsequently transform them into interpretable operations to realize personalized network resource allocations. Likewise, wireless communication operators can send commands to large AI models based on pre-defined functions, or they can utilize voice and other methods for real-time customization of instructions. When receiving these communication commands, the large AI models can activate corresponding functional modules, such as semantic communications and security threat prediction, to respond to the instructions, make smart decisions and relay them to network for execution. Indeed, these large AI models serve as coordinators within the wireless network, integrating the understanding and functionalities of sub-task models to offer a comprehensive, adaptive, and efficient wireless communication system. While existing AI research endeavors often suffered from high complexity and large number of optimization variables in large-scale heterogeneous networks, large AI models utilize huge computational resource to balance and optimize existing decision strategies, and seek optimal trade-offs among the network variables.

Due to the utilization of large language and visual models in future large AI models, the direct interaction between users and AI will elevate the personalization in wireless communications to a new height. Large AI models are not only proficient in understanding network parameters, but also adept at grasping the user’s context and potential needs. It has the capability to optimize network configurations, resource allocations, and service delivery based on individual user requirements to ensure that each user's wireless communication experience is optimized according to their specific needs and preferences. In addition, given the ever-changing environmental conditions in wireless communications, the real-time adaptability of large AI models proves to be highly advantageous. These AI models not only exhibit advanced decision-making capabilities upon pre-training, but also have the ability of continuous learning and adapting to new challenges and environments. The large AI models learn incessantly from real-time data and their decisions, utilizing network feedback to refine their decision-making strategies to ensure optimal transmission rates, minimized transmission delays and enhanced user experiences in wireless communications.

On the other hand, with the emergence of large AI models, generative AI has attracted much attention in AI research field, it demonstrates significant capabilities in understanding, generating and optimizing data. Thus, this primary generative capability makes generative AI as a powerful future solution for addressing the complex, dynamic and resource-constrained challenges within wireless communication systems. The learning capacity of large AI models for massive data can enhance the performance of generative AI, which in turn can offer data augmentation services to strengthen the performance of large AI models. Furthermore, considering scenarios such as edge networks, federated learning, and computation-limited nodes, generative models can be integrated with large AI models to provide personalized, distributed, and privacy-preserving AI training and deployment.

The integration of large AI models with future wireless communications is anticipated to herald a paradigm shift of the next communication age. These AI models are equipped to manage forthcoming large-scale wireless networks with high complexity and to ensure transmission efficiency, user personalization and security in wireless communications. However, as an emerging technology that has not been widely applied, large AI models will inevitably meet numerous significant challenges in future wireless communications. In the following section, we briefly summarize some potential challenges and corresponding solutions for large AI models in future wireless networks.

\section{Potential Challenges and Solutions of Large AI Model for Future Wireless Communications}\label{sec:3}
The integration of large AI models with wireless communications heralds a transformative era. However, this promising integration is not without challenges. Addressing the complexities associated with energy consumption, architectural design, privacy and security becomes paramount in future AI-driven wireless communications. In what follows, we discuss these challenges, exploring multifaceted issues and potential approaches for the continually evolving AI-driven wireless communications.

\subsection{Energy Consumption}
The impact of introducing large AI models in wireless communications are multifaceted, especially on energy consumption. The computational requirements for training the large AI models is very high, this is particularly pronounced in the training stage. Furthermore, the real-time data processing and decision-making of large AI models in dynamic network environments further contributing to the energy consumption. Moreover, the energy efficiency of large AI-driven wireless networks is intricately linked to their energy cost and environmental footprint. As the scale and complexity of AI models increase, the training cost, operational cost and carbon emission expands rapidly, this raises critical questions about the sustainability of large AI models, especially in scenarios where continuous learning and adaptation are essential.

Therefore, innovative algorithmic optimizations, energy-efficient hardware, and adaptive energy management protocols, will be essential in addressing the energy concerns for large AI models in wireless communications. A potential solution is that future wireless communication networks can employ large AI models to collect, analyze and predict data traffic across the entire network and edge networks to learn the optimal strategy for dynamic resource allocation to reduce the overall energy consumption. Besides, large AI-driven analysis of the network energy consumption can help develop novel energy management protocols to minimize unnecessary energy usage.

\subsection{Architecture design}
The integration of large AI models into wireless networks introduces a high level of complexity in architecture design. The large AI models require infrastructures which can deal with huge data processing and real-time analysis. The distributed AI model which operates at the network edge is a solution to meet the demand of personalized and real-time decision making. Such developments of large AI models need to address the balance problem of advanced computational processes while ensuring efficiency, coordination, and adaptability in future wireless networks. The evolution of large AI-driven wireless system towards more complex network structures aims to balance the huge demands posed by the integration of large AI models, and ensure that both computational and operational efficiency are satisfied and optimized.

However, this hybrid centralization/decentralization framework presents challenges associated with network coordination, data consistency and resource optimization, and significantly increases the complexity of future wireless network architectural design. As a solution, future research is estimated to focus on the development of hybrid architectures that integrate cloud, edge and fog computing to optimize data processing and decision-making in real-time. Moreover, network self-organization can become a key aspect in future wireless communications, especially in scenarios where AI models need to be deployed across various network nodes, each having different resource constraints and operational parameters.

\subsection{Privacy and Security}
Privacy and security are at the forefront of challenges in AI-driven wireless communications. The data-intensive nature of large AI models raises privacy concerns. Considering the fact that users’ sensitive data must be protected, the future AI-driven wireless communications require mechanisms that ensure data privacy while enabling the models to learn and adapt effectively. Current techniques, e.g. differential privacy and homomorphic encryption, were proposed as potential solutions to protect data privacy in federated learning frameworks, but introducing these techniques in dynamic wireless networks brings their own challenges. Therefore, future data privacy researches need to develop advanced cryptographic methods and privacy preserving algorithms to protect user data in large AI-driven wireless networks.

Moreover, security is another critical concern for future large AI-driven wireless networks. The integration of AI introduces vulnerabilities to new types of attacks, such as adversarial attacks which affect the AI training processes. Secure federated learning is deemed as a potential strategy to enhance security, it trains models collaboratively across multiple devices while keeping the data localized. However, it is a complex task for such distributed learning algorithms to ensure the integrity, authenticity, and confidentiality of data and models. Hence, the incorporation of more advanced security protocols and defensive algorithms will have paramount importance in future wireless communications. A potential solution is the intersection of AI and blockchain technology which can provide enhanced data integrity and security, and can be a promising frontier of future secure communications.

\subsection{Scalability}
Scalability of large AI models for future wireless communications is an important issue. As wireless networks develop, they are expected to support increasing devices, applications and services. Each addition introduces new dimensions of complexity and requires that AI models are architecturally equipped to adapt. Therefore, the large AI models need modular designs that allow for the integration of additional computational and learning modules without compromising the overall network performance. In addition, scalability is also about the economic viability of expanding AI-driven wireless networks. The cost implications of scaling large AI models are significant in computational resources, energy consumption and maintenance cost. Developing cost-effective scaling strategies that balance performance improvement with sustainability is a critical aspect of addressing scalability concerns.

To address this issue, the role of edge computing will be more important in addressing scalability of large AI models in future wireless communications. As networks expand, moving computational resources to the edge becomes a viable strategy to reduce latency and enhance real-time processing. Large AI models should be designed to leverage edge computing effectively, considering distributed computational loads and ensuring that data processing and decision-making could be localized to reduce the pressure on central resources and enhance the overall network efficiency. Moreover, edge computing architectures are adaptable for the seamless integration of new technologies and devices. This adaptability is very important to the scalability that the wireless network can evolve to meet emerging demands and challenges.

\subsection{Ethical and Regulatory}
Ethical and regulatory issues have attracted much attention in modern society. In future large-AI driven wireless communication networks, decisions made by AI may bring uncontrollable impacts, which leads to concerns of ethics. Therefore, it is important to build ethical frameworks that guide the decision-making processes and trainning methods of large AI to ensure alignment with modern societal values.

Moreover, with AI becoming an integral component of future wireless communications, regulatory institutions need to have in-depth understanding of legal, ethical, and technical domains in AI-driven wireless communications. Thus, the formulation and implementation of regulatory standards attuned to the era of large AI are imperative to protect users's security and privacy. As a solution, future AI researches need to align AI innovations with existing and future legal frameworks, such as international standards for data protection, cybersecurity and fair transmission.

\section{Future Directions}\label{sec:4}
As our society moves into an era of the integration of AI and wireless communications, the role of large AI models is becoming increasingly pivotal. However, it is essential to maintain a forward-looking perspective. Apart from addressing potential challenges, we also need to explore the prospective research directions associated with integrating large AI models into wireless communications, with the aim to shape the future of research in this area. This anticipatory approach provides an overall understanding and strategize perspective solutions, and underscore the progressive evolution of large AI-driven wireless communication networks.

\subsection{Adaptive Learning and Optimization Algorithms}
One of the main future directions for the integration of large AI models in wireless communications is the development of adaptive learning and optimization algorithms. As wireless networks become increasingly complex and dynamic, more new tasks and data will emerge, thus large AI models are confronted with the challenges of real-time learning and adaptation. In the future, these AI models need to be equipped with advanced algorithms capable of self learning, self optimization and self decision-making to ensure ongoing network optimization.

The adaptability of future large AI models will also need to be extended to security and privacy preservation. Future adaptive learning and optimization algorithms for large AI models will consider advanced security protocols and privacy-preserving algorithms to ensure that AI-driven decisions are not only optimal for transmission performance but are also secure and privacy-compliant. In essence, adaptive learning and optimization algorithms are the engines that will drive the future of AI-driven wireless communication systems. Therefore, the future adaptive AI models will be adaptive and efficient to ensure that the convergence of AI and wireless communications translates into networks that are not only connected but also intelligent, responsive and user-centric.

\subsection{Global Connectivity and Digital Inclusion}
The complexity of managing global connectivity is ever increasing due to the diverse and dynamic nature of user demands, environmental conditions and technological infrastructures. Large AI models are estimated to drive the realization of a globally connected world. With the integration of advanced technologies like satellite communications and low Earth orbit (LEO) satellite constellations, the large AI models are expected to address the challenges of ensuring ubiquitous connectivity. Large AI models will be utilized to support the global satellite communications with its analyzing ability, and ensure that satellite networks are optimized for performance and security, especially in remote and underserved areas.

On the other hand, digital inclusion plays an key role in future global connectivity framework. It not only ensures that connectivity is universal but also confirms the fairness for all users. The large AI models will be utilized in identifying gaps for digital access, in designing connectivity solutions to address these gaps and in ensuring that the future wireless networks is accessible to all users, regardless of geographical, economical, or social barriers.

\subsection{Custom AI Model Development}
As wireless communications continue to expand in terms of applications, users and tasks in future wireless communications, the one-size-fits-all approach is becoming increasingly insufficient. The emergence of custom AI models presents a resolution which offers customized solutions according to specific needs and challenges in varied wireless communication environments. They focus on particular applications, user analysis or network environments, embodying a blend of precision and flexibility. In large AI models, these custom models can help ensure a tailored and efficient response to the specific challenges from sub-tasks of wireless communications. Therefore, custom AI model development is a future direction that promises to redefine the landscape of large AI models in wireless communications.

\section{Conclusions}\label{sec:con}
In this paper, we introduced the background and importance of large AI models in future wireless communications. We further explored how these large AI models can amalgamate with existing AI applications to construct huge AI clusters to handle complex tasks in wireless networks. Addressing the prospective applications of large AI models in wireless communications, we described potential challenges encompassing security, privacy, scalability, architecture design, ethical, regulatory and energy consumption. Furthermore, we discussed the future research directions for large AI models in wireless communications. Overall, it becomes increasingly evident that large AI models are not only an indispensable element but a transformative force capable of redefining the structure of wireless communication networks.


\bibliographystyle{ieeetr}
\bibliography{ref}

\noindent\\
\textbf{Chong Huang} (Member, IEEE) received the Ph.D. degree in wireless communications from the University of Surrey in 2022. He is currently a Senior Research Fellow with the Institute for Communication Systems, 5GIC \& 6GIC, University of Surrey. His research interests include deep learning, cooperative networks, physical layer security, edge computing, reconfigurable intelligent surfaces, federated learning, satellite communications and Internet of Things.\\

\noindent
\textbf{Gaojie Chen} (Senior Member, IEEE) received the Ph.D. degree in electrical and electronic engineering from Loughborough University, Loughborough, U.K. He is currently a Professor with the School of Flexible Electronics (SoFE) \& State Key Laboratory of Optoelectronic Materials and Technologies, Sun Yat-sen University, China, and a Visiting Research Collaborator with the Information and Network Science Lab, University of Oxford.\\

\noindent
\textbf{Pei Xiao} (Senior Member, IEEE) received the Ph.D. degree from the Chalmers University of Technology, G\"oteborg, Sweden, in 2004. He is a Professor of Wireless Communications with the Institute for Communication Systems, home of 5GIC and 6GIC with the University of Surrey, Guildford, U.K. He is currently the Technical Manager of 5GIC/6GIC, leading the research team in the new physical-layer work area, and coordinating/supervising research activities across all the work areas.\\

\noindent
\textbf{Zhu Han} (Fellow, IEEE) currently is a professor in the Electrical and Computer Engineering Department at the University of Houston, Texas. He has been an AAAS Fellow since 2019. He received the IEEE Kiyo Tomiyasu Award in 2020. He has been a 1 percent highly cited researcher since 2017 according to Web of Science.\\

\noindent
\textbf{Rahim Tafazolli} (Senior Member, IEEE) is Regius Professor of Electronic Engineering at the University of Surrey, United Kingdom, and the director of the university’s Institute for Communication Systems and 5G Innovation Centre. He has been named a Fellow of the Royal Academy of Engineering in 2020, and appointed a Fellow of the Wireless World Research Forum in April 2011 in recognition of his personal contributions to the wireless world for heading one of Europe’s leading research groups.
\end{document}